\documentclass{icrc2009}

\usepackage{graphicx}   % for including figures
\usepackage{caption}    % for captions
\usepackage[font=footnotesize]{subfig} % subfig.sty for a double column floating figure using two subfigures 
\usepackage{fixltx2e}
\usepackage{url}

\newcommand{\shorttitle}[1]%
{\markboth{Proceedings of the 31\MakeLowercase{$^{st}$} ICRC, {\L}\'{o}d\'{z} 2009}{#1} }
\newcommand{\etal}{\MakeLowercase{\textit{et al. }}} % "et al."

\newcommand{\gc}{Galactic Centre}
\newcommand{\antares}{ANTARES}

\begin{document}
\title{Concepts and performance of the \antares\ data acquisition system}

\author{\IEEEauthorblockN{Mieke Bouwhuis\IEEEauthorrefmark{1}
		on behalf of the \antares\ collaboration}
\IEEEauthorblockA{\IEEEauthorrefmark{1}National Institute for Subatomic
Physics (Nikhef), Amsterdam, The Netherlands}}

% please write the preseter's name and short title (3-4 words maximum)
%    which will appear at the header of the even pages.
\shorttitle{Bouwhuis \etal \antares\ DAQ system}
\maketitle

\begin{abstract}

The data acquisition system of the \antares\ neutrino telescope
is based on the unique ``all-data-to-shore" concept.
In this, all signals from the photo-multiplier tubes are digitised,
and all digital data are sent to shore where they are processed in
real time by a PC farm.
This data acquisition system showed excellent stability and
flexibility since the detector became operational in March 2006.
The applied concept enables to operate different physics triggers to the
same data in parallel, each optimised for a specific (astro)physics
signal. 
The list of triggers includes two general purpose muon triggers, a 
\gc\ trigger, and a gamma-ray burst trigger.
The performance of the data acquisition system is evaluated by its
operational efficiency and the data filter capabilities.
In addition, the efficiencies of the different physics triggers are
quantified. 
\end{abstract}

 
\section{Introduction}

The \antares\ neutrino telescope is situated in the Mediterranean Sea
at a depth of about 2500~m, approximately 40~km south east of the
French town of Toulon.  
Neutrinos are detected through the detection of Cherenkov light
produced by the charged lepton that emerges from a neutrino
interaction in the vicinity of the detector. 
Measurements are focused mainly on muon-neutrinos, since the muon resulting from
a neutrino interaction can travel a distance of up to several kilometres.
Due to the transparency of the sea water (the absorption length is about
50~m), the faint Cherenkov light can be detected at relatively large
distances from the muon track. 
A large volume of sea water is turned into a neutrino
detector by deploying a 3-dimensional array of light sensors in the
water. 

The \antares\ detector consists of thirteen lines, each with up to 25~storeys.
The storeys are connected by cables which provide mechanical
strength, electrical contact and fibre-optic readout. 
Each line is held on the seabed by a dead-weight anchor and 
kept vertical by a buoy at the top of its 450~m length.
Along eleven lines, 25~storeys with three light sensors are placed 
at an inter-storey distance of 14.5~m starting 100~m above the seabed.
On each storey three spherical glass pressure vessels contain 10''
Hamamatsu photo-multiplier tubes (PMTs), which are oriented with their
axes pointing downward at an angle of 45~degrees from the vertical.  
One line consists of 20~such storeys, and one line is equipped with deep-sea
instrumentation.
Each storey in the detector has a titanium cylinder which houses the electronics for
data acquisition and slow control.
This system is referred to as the local control module.
In addition, each line has a line control module that is located at
the anchor. 

Daylight does not penetrate to the depth of the \antares\ site. 
Therefore the telescope can be operated day and night.
But even in the absence of daylight, a ubiquitous background
luminosity is present in the deep-sea due to the decay of radioactive
isotopes (mainly $^{40}$K) and to bioluminescence. 
This background luminosity produces a relatively high count rate of
random signals in the detector (60--150~kHz per PMT).
This background can be suppressed by applying the characteristic
time-position correlations that correspond to a passing muon to the
data. 

\section{Data acquisition system}

The main purpose of the data acquisition (DAQ) system is to convert
the analogue pulses from the PMTs into a readable input for the
off-line reconstruction software. 
The DAQ system is based on the ``all-data-to-shore" concept~\cite{daq}.
In this, all signals from the PMTs that pass a preset threshold
(typically 0.3 photo-electrons) are digitised and all digital data are
sent to shore where they are processed in real-time by a farm of
commodity PCs. 

\subsection{Network architecture}

The network architecture of the off-shore DAQ system has a star topology.
In this, the storeys in a line are organised into separate
sectors, each consisting of 5~storeys, and the detector lines are connected
to a main junction box. 
The junction box is connected to a station on shore via a single
electro-optical cable. 
The network consists of a pair of optical fibres for each detector line,
an 8~channel dense wavelength division \hbox{[de-]multiplexer} (DWDM)
in each line control module (200~GHz spacing), a small Ethernet switch
in each sector and a processor in each local control module.
The Ethernet switch in the sector consists of a combination of the
Allayer AL121 (eight 100~Mb/s ports) and the Allayer AL1022 (two Gb/s
ports). 
One of the 100~Mb/s ports is connected to the processor in the local
control module via its backplane (100Base--CX) and four are connected
to the other local control modules in the same sector via a
bi-directional optical fibre (100Base--SX). 
One of the two Gb/s ports is connected to a DWDM transceiver
(1000Base--CX). 
The DWDM transceiver is then 1--1 connected to an identical
transceiver on shore using two uni-directional optical fibres
(1000Base--LH). 
The line control module has also a processor which is connected to a
DWDM transceiver via its backplane (100Base--CX). 
A similar pair of DWDM transceivers is then used to establish a
100~Mb/s link to shore (100Base--LH). 
The network architecture on-shore consists of an optical
\hbox{[de-]multiplexer} and 6~DWDM transceivers for each detector
line, a large Ethernet switch (192~ports), a data processing farm and
a data storage facility. 
The optical fibres and the Ethernet switches form together a (large)
local area network. 
Hence, it is possible to route the data from any local control module
in the detector to any PC on shore. 

\subsection{Readout}

The front-end electronics consist of custom built analogue ring
sampler (ARS) chips which digitise the charge and the time of the
analogue signals from the PMT. 
The combined data is generally referred to as a hit; it can be a
single photo-electron hit or a complete waveform. 
The arrival time is determined from the signal of the clock system in
the local control module. 
An on-shore clock system (master) drives the clock systems in the
local control modules (slaves). 
The processor in the local control module is a Motorola MPC860P.
It runs the VxWorks real time operating system~\cite{vxworks} and
hosts the DaqHarness and ScHarness processes. 
The DaqHarness and ScHarness are used to handle respectively the data
from the ARS chips and the data from the various deep-sea instruments.
The latter is usually referred to as slow control data.
The processor in the local control module has a fast Ethernet
controller (100~Mb/s) that is connected to the Ethernet switch in the
sector. 
Inside the local control module, two serial ports with either RS485 or
RS232 links and the MODBUS protocol are used to handle the slow
control signals. 
The specific hardware for the readout of the ARS chips is implemented
in a high density field programmable gate array (XilinX Virtex-E
XCV1000E). 
The data are temporarily stored in a high capacity memory (64~MB
SDRAM) allowing a de-randomisation of the data flow.
In this, the data are stored as an array of hits.
The length of these arrays is determined by a predefined time frame of
about 100~ms and the singles rates of the PMTs. It amounts to about
60--200~kB. 
The data are read out from this memory by the DaqHarness process and
sent to shore. 
All data corresponding to the same time frame are sent to a single
data filter process in the on-shore data processing system.

The on-shore data processing system consists of about 50 PCs
running the Linux operating system. 
To make optimal use of the multi-core technology, four data filter
processes run on each PC. 
The physics events are filtered from the data by the data filter
process using a fast algorithm. 
The typical time needed to process 100~ms of raw data amounts to
500~ms. 
The available time is used to apply designated software triggers to
the same data. 
For example, a trigger that tracks the \gc\ is used whenever the count
rates of the PMTs are below 150~kHz (this corresponds to about 80\% of
the time). 
On average, the data flow is reduced by a factor of about 10,000.
The filtered data are written to disk in ROOT format~\cite{ROOT} by a central data
writing process. 
The count rate information of every PMT is stored together with the
physics data. 
The sampling frequency of these rate measurements is about 10~Hz.
The data from the readout of the various instruments are transferred
as an array of parameter values and stored in the database via a
single process. 
The readout of the various deep-sea instruments is scheduled via read
requests that are sent from shore by a designated process. 
The frequency of these read requests is defined in the database.
A general purpose data server based on the tagged data concept is used
to route messages and data~\cite{controlhost}. 
For instance, there is one such server to route the physics
events to the data writer which is also used for online monitoring.
Messages (warnings, errors, etc.) are collected, displayed and written
to disk by a designated GUI. 

\subsection{Operation}

The main control GUI allows the operator to modify the state of the
system. 
In total, the system involves about 750~processes (300~DaqHarness
processes, 300~ScHarness processes, 120~data filters, and various other
processes).
These processes implement the same state machine diagram~\cite{chsm}.
Before the start of a data taking run, the whole system including the
detector is configured. 
In order to archive data efficiently, the main control GUI updates
the run number regularly. 
The database system is used to keep track of the history of the
detector and the data taking. 
It is also used for storing and retrieving configuration parameters of
the whole system. 
The positions of the PMTs are determined using a system of acoustic
transmitters and receivers. 
The corresponding data are recorded at the same time as the physics
data.
The time calibration of the PMTs is obtained using special data taking
runs. 
During these runs, one or more LED beacons (or laser beacon) flash.
The typical flash rate is about 1~kHz.
The time calibration data are recorded using a designated software
trigger. 
All data are archived in the IN2P3 computer centre in Lyon which also
houses the Oracle database system. 

\subsection{External triggers}

The on-shore data processing system is linked to the gamma-ray bursts
coordinates network (GCN)~\cite{GCN}. 
This network includes the Swift and Fermi satellites.
There are about 1--2 GCN alerts per day and half of them correspond to
a real gamma-ray burst. 
For each alert, all raw data are saved to disk during a preset period
(presently 2 minutes). 
The buffering of the data in the data filter processors is used to
save the data up to about one minute before the actual alert~\cite{antares-grb}.

\section{Performance of the DAQ system}  

The performance of the DAQ system can be summarised in terms of the
efficiency to detect neutrinos and the efficiency to operate the
neutrino telescope. 
The efficiency to detect neutrinos is primarily determined by the
capability to filter the physics events from the continuous data
streams. 
With the all-data-to-shore system, different software triggers can be
operated simultaneously. 
At present, two general purpose muon triggers (`standard') and one
minimum bias trigger are used to take data. 
The minimum bias trigger is used for monitoring the data quality.
The standard trigger makes use of the general causality relation:

\begin{eqnarray}
   \left| t_i - t_j \right| \le \left| \bar{x}_i - \bar{x}_j \right|  \times \frac{n}{c} \label{eq:3D}, 
\end{eqnarray}

\noindent
where $t_i$ ($\bar{x}_i$) refers to the time (position) of hit $i$,
$c$ the speed of light and $n$ the index of refraction of the sea water.
In this, the direction of the muon, and hence the neutrino, is not used.
The standard trigger is therefore sensitive to muons covering the full
sky. 
To limit the rate of accidental correlations, the hits have been
preselected. 
This pre-selection (L1) includes coincidences between two neighbouring
PMTs in the same storey and large pulses (number of photo-electrons
typically greater than 3). 
The minimum number of detected photons to trigger an event ranges
between 4--5 L1 hits, depending on the trigger algorithm.
This corresponds to a typical threshold of several 100~GeV.
The purity of the trigger (fraction of events that correspond to a
genuine muon) has been determined using a simulation of the detector
response to muons traversing the detector and a simulation based on
the observed background. 
The purity is found to be better than 90\%.
The 10\% impurity is mainly due to (low-energy) muons that in
combination with the random background produce a trigger.
A small fraction of the events ($\ll 1\%$) is due to accidental
correlations. 
The observed trigger rate is thus dominated by the background of
atmospheric muons and amounts to \mbox{5--10~Hz} (depending on the
trigger conditions). 

\begin{figure}[t!]
\setlength{\unitlength}{1cm}
\begin{center}
\begin{picture}(8,8)
\put(0,0){\scalebox{0.45}{\includegraphics{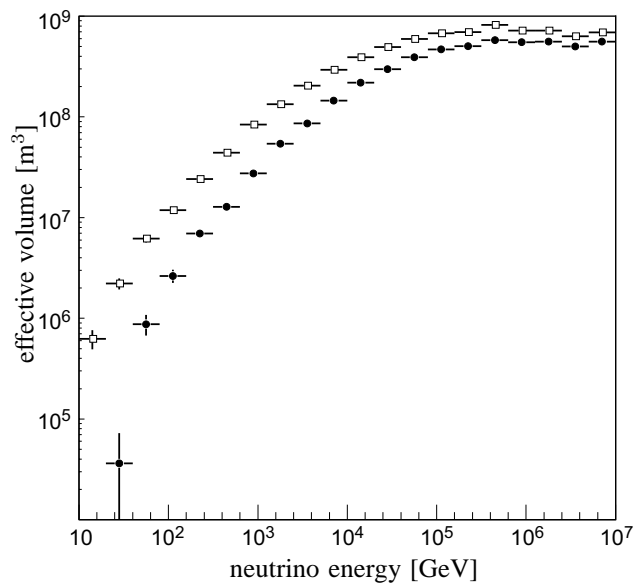}}}%
\put(4.5,0){\makebox(0,0)[b]{neutrino energy [GeV]}}%
\put(0,4.5){\makebox(0,0)[l]{\rotatebox{90}{effective volume [m$^3$]}}}%
\end{picture}
\end{center}
\caption{
Effective volume of the standard trigger (solid circles) and the \gc\
trigger (open squares) as a function of the neutrino energy. 
\label{f:volume}
}
\end{figure}

In addition to the standard trigger, a trigger that tracks the \gc\ is
used to maximise the detection efficiency of neutrinos coming from the
\gc. 
This trigger makes use of the direction specific causality relation:

\begin{eqnarray}
   (z_i - z_j) - R_{ij}\tan{\theta_C} \le c(t_i - t_j) \nonumber \\
  ~~~~~~~~~\le (z_i - z_j) + R_{ij}\tan{\theta_C} \label{eq:1D},
\end{eqnarray}

\noindent
where $z_i$ refers to the position of hit $i$ along the neutrino
direction, $R_{ij}$ refers to distance between the positions of hits
$i$ and $j$ in the plane perpendicular to the neutrino direction and
$\theta_C$ to the characteristic Cherenkov angle. 
Compared to equation \ref{eq:3D}, this condition is more stringent
because the 2-dimensional distance is always smaller than the
3-dimensional distance. 
Further more, this distance corresponds to the distance travelled by
the photon (and not by the muon). 
Hence, it can be limited to several absorption lengths (e.g.\ 100~m or
so) without loss of detection efficiency. 
This restriction reduces the combinatorics significantly (about a
factor of 10 for each additional hit). 
As a consequence, all hits can be considered and not only preselected
hits (L1) without compromising the purity of the physics events. 
The detection efficiency of the general purpose muon trigger and the
\gc\ trigger are shown in Fig.~\ref{f:volume}. 
The effective volume is defined as the volume in which an interaction
of a muon neutrino would produce a detectable event.
\noindent
As can be seen from Fig.~\ref{f:volume}, the detection efficiency
obtained with the \gc\ trigger is significantly higher than that
obtained with the standard trigger. 
As a result, the sensitivity to neutrinos from the \gc\ is
greatly improved. 
The field of view for which this improved efficiency is obtained is about 
10~degrees.

Since the start of the operation of the detector (March 2006), the
average data taking efficiency has been better than 90\%.

\enlargethispage{-1.0in}
\end{document}